*Research Article*

# Phytochemical Study and Evaluation of the Cytotoxic Properties of Methanolic Extract from *Baccharis obtusifolia*

**Juan Carlos Romero-Benavides**,[1] **Gina C. Ortega-Torres**,[1] **Javier Villacis**,[2] **Sara L. Vivanco-Jaramillo**,[1] **Karla I. Galarza-Urgilés**,[1] and **Natalia Bailon-Moscoso**[2]

[1]*Departamento de Química y Ciencias Exactas, Universidad Técnica Particular de Loja (UTPL), San Cayetano Alto S/N, 1101608 Loja, Ecuador*
[2]*Departamento de Ciencias de la Salud, Universidad Técnica Particular de Loja (UTPL), San Cayetano Alto S/N, 1101608 Loja, Ecuador*

Correspondence should be addressed to Juan Carlos Romero-Benavides; jcromerob@utpl.edu.ec

Some species of the *Baccharis* genus have been shown to possess important biomedical properties, including cytotoxic activity. In this study, we examined the cytotoxic effect of methanol extract from *Baccharis obtusifolia* (Asteraceae) in cancer cell lines of prostate (PC-3), colon (RKO), astrocytoma (D-384), and breast (MCF-7). The methanolic extract displayed the largest substantial cytotoxic effect in lines of colon cancer (RKO) and cerebral astrocytoma (D-384). Chromatographic purification of the *B. obtusifolia* methanolic extract led to the isolation and identification of 5,4′-dihydroxy-7-methoxyflavone (**1**) and 5-hydroxy-7,4′-dimethoxyflavone (**2**) compounds of the flavonoid type.

## 1. Introduction

*Baccharis* is the largest genus in the family Asteraceae, with over 500 species distributed throughout North and South America [1]. The largest American genus *Baccharis* (Asteraceae) includes about 400 species. Of these species, 20% are locally used for medical purposes or, to a lesser extent, as food or as raw material for different local industries [2]. The most prominent compounds in *Baccharis* are diterpenoids, phenolic compounds like flavonoids and coumarins, and triterpenoids, among others. Flavonoids are important compounds isolated from a wide range of plants [3]. Diets with a high flavonoid content are associated with positive health effects and the prevention of several diseases [4]. Additionally, pharmacological studies have demonstrated the anti-inflammatory effects [3] and antioxidant capacity [5] of several flavonoids, and some flavonoids have been demonstrated to possess cytotoxic, antifungal, antiviral, and antibacterial properties [5–7]. The most prominent biomedical applications of the *Baccharis obtusifolia* H.B.K., commonly known as "Chilca redonda," include the treatment of rheumatism, liver disease, wounds, and ulcers [1]. In the present study, we performed a chemical composition analysis of the active phytometabolites obtained from the methanolic extract of the leaves of *Baccharis obtusifolia*. We also evaluated the cytotoxic activity of this extract on human cancer cell lines.

## 2. Materials and Methods

*2.1. Preparation of the Extracts.* The leaves from the *Baccharis obtusifolia* species were collected on Villonaco (04°01′25″ Lat. S, 79°14′45″ Long. O, 2849 m.a.s.l.) mountain of the Loja Province of Ecuador. A sample specimen (PPN-as-014) was deposited and identified in the Herbarium of Departamento de Química of Universidad Técnica Particular of Loja, Ecuador.

The collected leaves were subjected to a dehydration process in a drying tray with airflow at a temperature of 32°C for seven days (final humidity: 6.8%).

To obtain the extract, we used 145 g of dried leaves and cold methanol (4-5°C). The method employed was dynamic maceration for 5 hours in a light-free environment. This

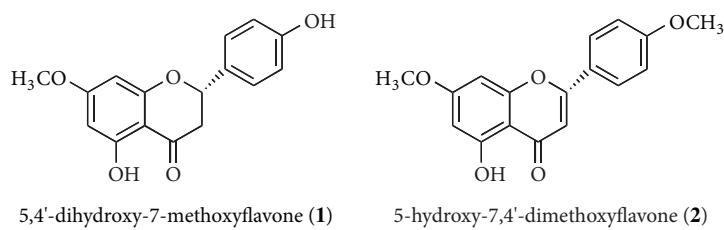

5,4'-dihydroxy-7-methoxyflavone (**1**)     5-hydroxy-7,4'-dimethoxyflavone (**2**)

Figure 1: Structure of compounds isolated from *Baccharis obtusifolia*.

procedure was repeated three times. The extract was then concentrated at 50 mbar and 37°C on a rotary evaporator (Buchi R210; Switzerland, Flawil) and stored at 4°C.

### 2.2. Isolation of Secondary Metabolites.
The methanolic extract (8.74 g) was filtrated to remove chlorophylls with a reverse phase silica gel RP-18, 6.5 cm (CC, Merck, Darmstadt, Germany) using specific mixtures of solvents: fraction 1 MeOH/$H_2O$ 85:15 (1000 ml), fraction 2 MeOH/$H_2O$ 90:10 (250 ml), fraction 3 100% MeOH (500 ml), and fraction 4 100% dichloromethane (500 ml).

To separate the components from the extract without chlorophylls (fraction 1), a silica gel-60 F254 chromatography column (CC, Merck) was used. Mixtures of hexane and ethyl acetate solvents were used in polarity starting with hexane (100%) to separate the compounds. Fractions of 200 ml each were collected using a vacuum pump, and the solvent was then removed on a rotary evaporator and the residue was recovered with dichloromethane. Thin layer chromatography (TLC, Merck) was performed on each fraction to detect the compounds. The compounds were visualized by spraying the solutions with a mixture of vanillin, ethanol, and acid sulfuric followed by heating on a hot plate. Fractions with a similar profile were pooled and purified by conventional procedures. Fraction 40-43 was crystallized using hexane and ethyl acetate and fraction 70-79 was crystallized using hexane and petroleum ether.

### 2.3. Characterization and Identification of Secondary Metabolites.
We followed the methods of Bailon-Moscoso et al. (2016) [8]. Melting points were determined using a Fisher-Johns apparatus. The $^1H$ and $^{13}C$ NMR spectra were recorded at 400 MHz and 100 MHz, respectively, on Varian 400 MHz-Premium Shielded equipment (Varian, Massachusetts, USA) using tetramethylsilane as an internal reference. $CDCl_3$ and DMSO-d6 were used as solvents; chemical shifts were expressed in parts per million (ppm) and coupling constants (*J*) were reported in Hz. Mass spectra (MS) were determined by a gas chromatograph (Agilent Technologies 6890 N, Wilmington, DE) coupled to a mass spectrometer (Agilent Technologies 5973 inert).

### 2.4. Cell Culture Procedures.
Four human cancer cell lines were used: PC-3, RKO, D-384, and MCF-7. The cells were cultured in RPMI-1640 medium supplemented with 10% fetal bovine serum (FBS, Invitrogen, Karlsruhe, Germany), 1% antibiotic-antimitotic solution (100 units/ml penicillin G, 100 $\mu$g/ml streptomycin, and 0.25 $\mu$g/ml amphotericin B, Gibco, Grand Island, NY), and 1% L-glutamine (2 mM, Gibco). The cells were incubated at 37°C in a 5% $CO_2$ atmosphere. The viable cells were counted using the trypan blue exclusion method in a hemocytometer [8].

### 2.5. Cell Viability Analysis via MTS Assay.
The MTS (5-[3-(carboxymethoxy)phenyl]-3-(4,5-dimethyl-2-thiazolyl)-2-(4-sulfo-phenyl)-2H-tetrazolium inner salt) cell viability assay was used to assess the inhibitory effects of the extracts on the survival of human cancer cell lines. A total of 3–5 $\times 10^3$ cells/well were seeded into 96-well plates and were allowed to adhere for 24 h. The cells were then treated with 50 $\mu$g/ml of whole extract to yield a final volume of 2 ml. Each concentration/assay was performed three times. Dimethyl sulfoxide (DMSO) was used as a negative control at a final concentration of 0.1% v/v, and 1 $\mu$M doxorubicin was used as a positive control. The cells were incubated with the treatments for 48 h, after which 20 $\mu$L MTS (5 mg/ml, Aqueous One Solution Reagent, Gibco) was added and further incubated for 4 h at 37°C. The absorbance was measured at 570 nm. The data obtained with cells treated with DMSO were considered to represent 100% viability [8]. In cell lines with an inhibition percentage over 30%, doses of 15, 45, 60, 75, and 100 $\mu$g/ml of methanolic extract were applied, and the MTS assay was used to measure the inhibitory concentration 50 ($IC_{50}$) using nonlinear regression.

### 2.6. Statistical Analysis.
All of the data were reported as means ± *standard error of the mean* (SEM) of three independent experiments. The nonlinear regression was determined by GraphPad Prism 5 (GraphPad Software, San Diego, CA).

## 3. Results

### 3.1. Isolation of Secondary Metabolites from B. Obtusifolia.
From the methanolic extract without chlorophylls, two representative fractions were obtained: F 70-79 eluted in Hex:EtOAc 80:20, from which we isolated 80 mg of a white crystalline solid identified as sakuranetin (5,4'-dihydroxy-7-methoxyflavanone) (**1**), and the fraction F 43-45 eluted in Hex:EtOAc 90:10, from which we isolated 11.2 mg of a yellow crystalline solid identified as 5-hydroxy-7,4'-dimethoxyflavone (**2**) (Figure 1). All of the compounds were identified based on physical and spectroscopic data via comparisons with the literature [9–11].

Table 1: Effect of *Baccharis obtusifolia* extract on the growth of human cancer cell lines.

| Treatment | % of inhibition ± SEM[a] | | | |
| --- | --- | --- | --- | --- |
| | Human cancer cell lines | | | |
| | MCF-7 (breast cancer) | PC3 (prostate cancer) | RKO (colon cancer) | D-384 (astrocytoma) |
| Methanolic extract | 20.2 ± 0.7 | 18.3 ± 1.2 | 89.2 ± 1.3 | 60.6 ± 0.5 |
| Doxorubicin (1 $\mu$M) | 40.9 ± 7.4 | 54.8 ± 1.9 | 89.2 ± 4.0 | 79.0 ± 4.6 |

[a]Mean and standard error (SEM) of at least three independent experiments. Control cells were considered to exhibit 100% viability.

Table 2: Half maximal inhibitory concentration ($IC_{50}$) of the compounds isolated from *Baccharis obtusifolia* on cancer cell lines.

| Compound | $IC_{50}$ ± SEM[a] ($\mu$M) | | | |
| --- | --- | --- | --- | --- |
| | MCF-7 (breast cancer) | PC-3 (prostate cancer) | RKO (colon cancer) | D-384 (astrocytoma) |
| 5,4′-dihydroxy-7-methoxyflavone (**1**) | >100 | >100 | 68.11 ± 1.08 | >100 |
| 5-hydroxy-7,4′-dimethoxyflavone (**2**) | >100 | >100 | 34.30 ± 1.34 | >100 |

[a]Each datum is given as the mean and its standard error (SEM) of at least three independent experiments.

*3.2. Characterization and Identification of Secondary Metabolites.* Physical and spectroscopic constants from 5,4′-dihydroxy-7-methoxyflavone (**1**): m.p. 240–243°C, EIMS m/z (%) 286 (M$^+$, 100), 269 (36), 241 (16), 167 (13), 166 (14), 138 (14), 128 (16), 121 (7), 118 (10). $^1$H NMR (DMSO-d$_6$) $\delta$ ppm: 2.8 (1H, dd), 3.1 (1H, dd), 3.89 (3H, s), 6.40 (1H, d, $J$=2.3 Hz), 6.79 (1H, d, $J$=2.3 Hz), 6.85 (1H, s), 6.96 (2H, d, $J$=9.3 Hz), 7.98 (2H, d, $J$=9.3 Hz), 12.95 (1H, s,) $^{13}$C NMR (DMSO-d$_6$) $\delta$ ppm: 79 (C-2) 43.1 (C-3) 196.0 (C-4) 164.1 (C-5) 95.1 (C-6) 168.0 (C-7) 94.2 (C-8) 162.8 (C-9) 103.5 (C-10) 130.5 (C-1') 128.0 (C-2') 114.2 (C-3') 156.1 (C-4') 115.6 (C-5') 128.0 (C-6') 55.6 (OCH$_3$-7).

Physical and spectroscopic constants from 5-hydroxy-7,4′-dimethoxyflavone (**2**): m.p. 160–163°C. EIMS m/z (%) 298 (M$^+$, 100), 269 (28), 255 (14), 166 (11), 138 (16), 135 (10), 132 (16). $^1$H NMR (CDCl$_3$) $\delta$ ppm: 3.89, 3.90 (3H each, s), 6.37 (1H, d, $J$= 2.4 Hz), 6.49 (1H, d, $J$=2.4 Hz), 6.58 (1H, s), 7.02 (2H, d, $J$=9.3 Hz), 7.85 (2H, d, $J$=9.3 Hz), 12.80 (1H, s). $^{13}$C NMR (CDCl$_3$) $\delta$ ppm: 164.0 (C-2) 104.3 (C-3) 182.4 (C-4) 165.4 (C-5) 98.0 (C-6) 165.4 (C-7) 92.6 (C-8) 162.1 (C-9) 104.3 (C-10) 123.5 (C-1') 128.0 (C-2') 114.4 (C-3') 157.6 (C-4') 114.5 (C-5') 128.3 (C-6') 55.6 (OCH$_3$ 7') 55.7 (OCH$_3$-4').

*3.3. Cytotoxic Effects of Methanolic Extract and Isolated Compounds on Human Cancer Cell Lines.* The effect of the methanolic extract on cell viability was determined via an MTS assay using human cancer cell lines treated for 48 h with the whole extract of *B. obtusifolia* (50 $\mu$g/ml). The extract exhibited a strong inhibitory effect on RKO 89.2%, as in D-384 cells with 60.6%. However, the MCF-7 and PC3 cell lines exhibited moderate inhibitory activity with 20.2% and 18.3%, respectively (Table 1).

Compounds **1** and **2** presented higher $IC_{50}$ to 100 $\mu$M for MCF-7, D-384, and PC-3 cell lines; for the RKO cell line, we obtained $IC_{50}$ of 68.11 $\mu$M (19.48 $\mu$g/ml) and 34.30 $\mu$M (10.22 $\mu$g/ml), for compounds **1** and **2**, respectively. Compounds **1** and **2** were specific for the RKO line, and compound **2** was more effective than compound **1** (Table 2).

## 4. Discussion

The species belonging to the Asteraceae family are characterized by their abundant biological activity [12], which results from their chemical composition, including secondary metabolites such as terpenes, diterpenes, flavonoids, and coumarins [13]. In the present study, two flavonoid compounds were isolated, 5,4′-dihydroxy-7-methoxyflavanone (**1**) and 5-hydroxy-7,4′-dimethoxyflavone (**2**). These compounds have been identified for the first time in *Baccharis obtusifolia*, but they have been reported in other species: compound **1** in *Chromolaena subscandens* [14], *Trixis vauthieri* [15], and *Trigona spinipes* [16] and compound **2** in *Baccharis polycephala* [4], *Thymus vulgaris* [9], *Combretum erythrophyllum* [17], *Kaempferia parviflora* [18], and *Boesenbergia pandurata* [19].

Our evaluation of the biological activity of methanolic extract revealed a high percentage of inhibition in RKO cell lines (89.2%), followed by D-384 cells (60.6%). This activity can be attributed to the presence of flavonoids in the extract. Some flavonoids act on these stages in very different cell types, which may lead to the use of these compounds as cytostatic agents in the later stages of carcinogenesis rather than as early-stage preventive elements [20, 21].

$IC_{50}$ of flavonoids **1** and **2** shows similar values to the other compounds of the same type [22–24]. Numerous studies have suggested that flavonoids may play a protective role in the prevention of cancer, coronary heart diseases, bone loss, and many other age-related diseases [21]. The activity found in the methanol extract may be due to the presence of isolated flavonoids.

Chemically, flavones are substances of phenolic nature and are characterized by having two benzene aromatic rings united by a bridge with three carbon atoms. They have the general structure $C_6$-$C_3$-$C_6$, which may form a third ring. Natural flavonoids are often present in at least three phenolic hydroxyls [25]. The NMR $^1$H data from compounds **1** and **2** show a substitution pattern *"para"* in the ring B [26];

substitutions in these flavones are found in the C-4′ position of the B ring; in the case of compound **1**, the substituent corresponds to a hydroxyl group (OH), and for compound **2**, the substituent corresponds to a methoxy group (OCH$_3$).

In the NMR data of $^{13}$C from compound **1** in the aliphatic zone $\delta$ (55.6 ppm), there appears a signal that belongs to a methoxy group. The signal in $\delta$ (43.1 ppm) is assigned to C-3 [26] and the signal in 103.5 ppm is assigned to C-10 of hydroxylated flavones, which are included in the structure of our compound and confirmed in the $^1$H NMR spectrum. The spectral data of compound **2** show two signals in $\delta$ (55.6–55.7 ppm) belonging to two methoxy groups present in this structure, and these two groups are also observed in the $^1$H NMR spectrum.

Flavonoids have characteristics that make them attractive for anticancer research. They act *in vitro* via various mechanisms in the oncogenic process, which renders them possible useful agents in the early stages of cancer or in the inhibition of later stages of progression or invasion. This activity may be explained by studying the Structure-Activity Relationship (SAR) between several flavonoids and cancer [27]. An example is the study of Zhang (2005) which focused on activity of flavonoids and breast cancer cells. This author found that the double bond between C-2 and C-3, the ring B connected to C-2, the hydroxyl group in C-5, the nonhydroxylation in C-3, and the presence of apolar substituents in C-6, C-7, C-8, or C-4′ were structural characteristics important for the interaction between flavonoids and *breast cancer resistance protein* (BCRP) [21]. The structure 5-hydroxy-7,4′-dimethoxyflavone (**2**) possesses important structural characteristics like the double bond between C-2 and C-3 and the presence of an OH group at the 5-position and the carbonyl group at the 4-position. Previous studies have suggested that the position, number, and substitution of hydroxyl groups in rings A and B, as well as the saturation in the C2-C3 bond, may be important factors that increase the cytotoxic or antiproliferative activities of flavonoids [28], which would explain the cytotoxic action of 5-hydroxy-7,4′-dimethoxyflavone in RKO cells. On the contrary, the flavone sakuranetin (**1**) does not exhibit the double bond at C-2 and C-3, which in mono- or dihydroxylated flavones results in the loss of cellular growth inhibitory activity [28]. It would be interesting to continue studies of this species and the isolated compounds. Additional studies of the molecular mechanisms underlying the effect of these secondary metabolites on cancer cell survival are accordingly necessary.

## Data Availability

The dataset supporting this article is included in the manuscript.

## Disclosure

The authors alone are responsible for the content and writing of the paper.

## Conflicts of Interest

The authors report no conflicts of interest.

## Acknowledgments

This work was partially supported by Universidad Técnica Particular de Loja, Ecuador (Grant PROY_FIN_QUI_0008).

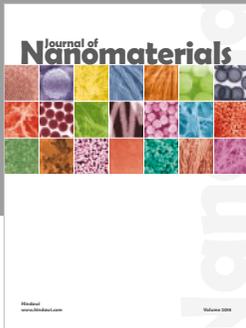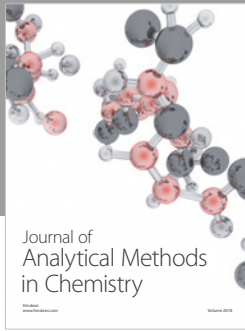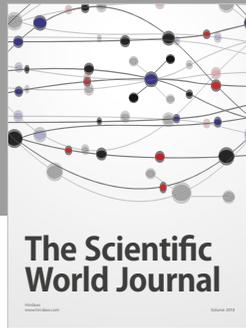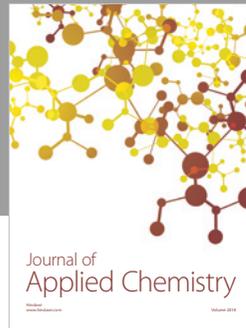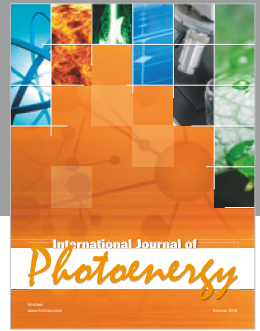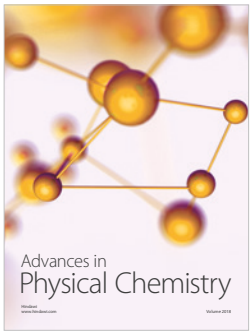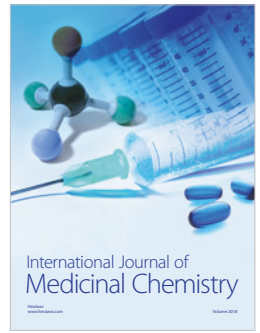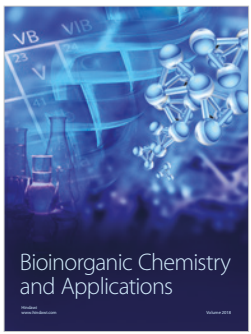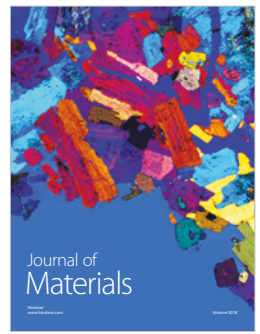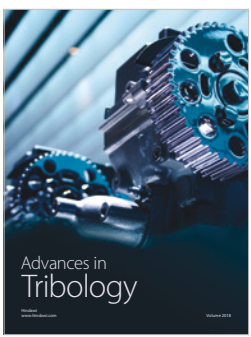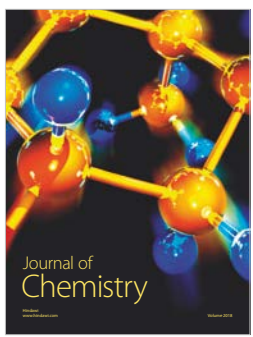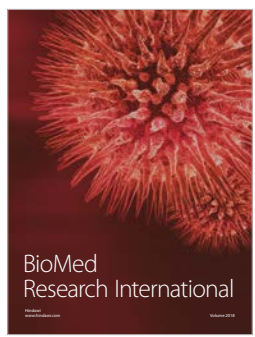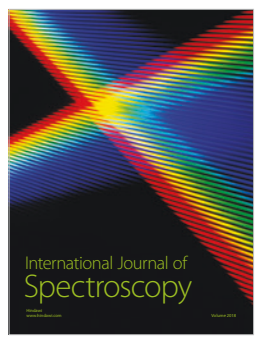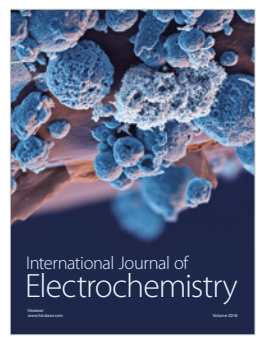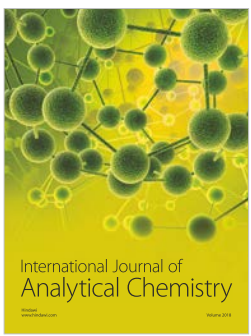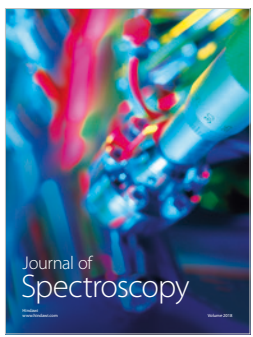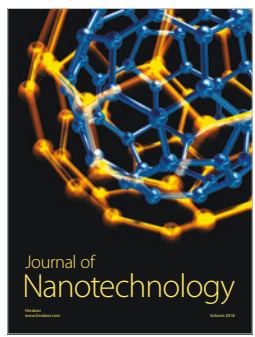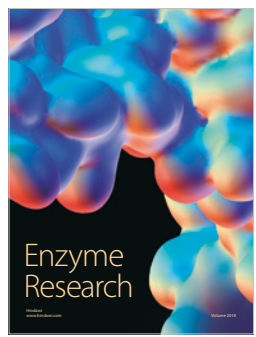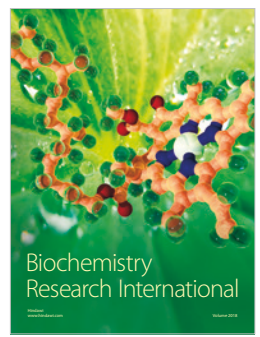